\documentclass[a4paper,12pt]{article}

\catcode`\@=11
\def\@eqnnum{{\rm [\theequation]}}
\def\@cite#1#2{({#1\if@tempswa , #2\fi})}
\catcode`\@=12

\usepackage{umlaut}
\usepackage{ifthen}
\usepackage[german,english]{babel}
\usepackage[dvips]{graphicx}
\usepackage{subfigure}

\newlength{\StdLen}
\newlength{\len}

\newcounter{celle}

\newcommand{\ChemZelle}[2]{%
   \def\_{\:}%
   \def\~{\,}%
   \refstepcounter{celle}%
   \def\Nr{\arabic{celle}}%
   \def\StdZelle{C \Nr}%
   \ifthenelse{\equal{#2}{}}{}{\renewcommand{\theequation}{\mbox{#2}}}%
   \begin{equation}\rm #1\end{equation}%
   \addtocounter{equation}{-1}\ifthenelse{\equal{#2}{}}{}%
      {\renewcommand{\theequation}{\arabic{equation}}}}

\newcommand{\ChemGl}[6]{%
   \def\_{\:}%
   \def\~{\,}%
   \def\Nr{\arabic{equation}}%
   \setlength{\StdLen}{1.3cm}%
   \setlength{\len}{#4}%
   \setlength{\unitlength}{\len}%
   \ifthenelse{\equal{#6}{}}{}{\renewcommand{\theequation}{\mbox{#6}}}%
   \begin{equation}
      \rm #1\;\;\;
      \ifthenelse{\equal{#5}{R}}{%
      \raisebox{0.8ex}%
         {\mbox{\beginpicture
                \setcoordinatesystem
                \put {\vector(1,0){1}} [Bl] at 0 0
                \endpicture}%
          \hspace{-1\len}%
          \raisebox{1ex}{\makebox[\len]{$\scriptstyle\rm #3$}}}}{%
      \ifthenelse{\equal{#5}{L}}{%
      \raisebox{0.8ex}%
         {\mbox{\beginpicture
                \setcoordinatesystem
                \put {\vector(-1,0){1}} [Bl] at 0 0
                \endpicture}%
          \hspace{-1\len}%
          \raisebox{1ex}{\makebox[\len]{$\scriptstyle\rm #3$}}}}{%
      \raisebox{1.2ex}%
         {\mbox{\beginpicture
                \setcoordinatesystem units <\len,1mm>
                \put {\vector(1,0){1}} [Bl] at 0 0
                \put {\vector(-1,0){1}} [Bl] at 1 -2
                \endpicture}%
          \hspace{-1\len}%
          \raisebox{1ex}{\makebox[\len]{$\scriptstyle\rm #3$}}}}}
      \;\;\;#2
   \end{equation}%
   \ifthenelse{\equal{#6}{}}{}{%
      \renewcommand{\theequation}{\arabic{equation}}}}

\newcommand{\Chemgl}[6]{%
   \def\_{\:}%
   \def\~{\,}%
   \def\Nr{\arabic{equation}}%
   \setlength{\StdLen}{1.3cm}%
   \setlength{\len}{#4}%
   \setlength{\unitlength}{\len}%
   \ifthenelse{\equal{#6}{}}{}{\renewcommand{\theequation}{\mbox{#6}}}%
   \[
      \rm #1\;\;\;
      \ifthenelse{\equal{#5}{R}}{%
      \raisebox{0.8ex}%
         {\mbox{\beginpicture
                \setcoordinatesystem
                \put {\vector(1,0){1}} [Bl] at 0 0
                \endpicture}%
          \hspace{-1\len}%
          \raisebox{1ex}{\makebox[\len]{$\scriptstyle\rm #3$}}}}{%
      \ifthenelse{\equal{#5}{L}}{%
      \raisebox{0.8ex}%
         {\mbox{\beginpicture
                \setcoordinatesystem
                \put {\vector(-1,0){1}} [Bl] at 0 0
                \endpicture}%
          \hspace{-1\len}%
          \raisebox{1ex}{\makebox[\len]{$\scriptstyle\rm #3$}}}}{%
      \raisebox{1.2ex}%
         {\mbox{\beginpicture
                \setcoordinatesystem units <\len,1mm>
                \put {\vector(1,0){1}} [Bl] at 0 0
                \put {\vector(-1,0){1}} [Bl] at 1 -2
                \endpicture}%
          \hspace{-1\len}%
          \raisebox{1ex}{\makebox[\len]{$\scriptstyle\rm #3$}}}}}
      \;\;\;#2
   \]%
   \ifthenelse{\equal{#6}{}}{}{%
      \renewcommand{\theequation}{\arabic{equation}}}}

\newcommand{\CaptionTab}[3]{\refstepcounter{table}%
   \addcontentsline{lot}{table}{\protect%
   \numberline{\thetable}{#2}}\label{#1}%
   \renewcommand{\baselinestretch}{1.2}\small\normalsize%
   \hspace*{\fill}\begin{minipage}{0.93\textwidth}{\hspace*{-0.07\textwidth}%
    \bf Tab. \thetable. }#3\end{minipage}\vspace{3mm}%
   \renewcommand{\baselinestretch}{1.0}\small\normalsize}
\newcommand{\CaptionFig}[3]{\refstepcounter{figure}%
   \addcontentsline{lof}{figure}{\protect%
   \numberline{\thefigure}{#2}}\label{#1}%
   \renewcommand{\baselinestretch}{1.2}\small\normalsize\rule{1mm}{0mm}\\[3mm]%
   \hspace*{\fill}\begin{minipage}{0.93\textwidth}{\hspace*{-0.07\textwidth}%
    \bf Fig. \thefigure. }#3\end{minipage}%
   \renewcommand{\baselinestretch}{1.2}\small\normalsize}

\renewcommand{\thesubfigure}{{\bf\Alph{subfigure}}}
\makeatletter
\renewcommand{\@thesubfigure}{\thesubfigure}
\makeatother
\newcommand{\uc}[1]{\uppercase{#1}}
\newcommand{\chap}[1]{\vspace{5mm} \centerline{\uc{\bf #1}} %
                      \vspace{3mm}}

\def\thebibliography#1{%
 \chap{\refname}%
 \list{\arabic{enumi}.}{\settowidth\labelwidth{[#1]}\leftmargin\labelwidth
 \advance\leftmargin\labelsep
 \usecounter{enumi}}
 \def\newblock{\hskip .11em plus .33em minus .07em}
 \sloppy\clubpenalty4000\widowpenalty4000
 \sfcode`\.=1000\relax}

\renewcommand{\baselinestretch}{1.2}

\begin{document}
\centerline{\large\bf Instrument Control at the FRM-II using TACO and NICOS}

\vspace{1cm}
\begin{center}
T. Unruh\\[2mm]
ZWE FRM--II\\
Technische Universität München\\
Lichtenbergstr. 1\\
D--85747 Garching, F.R.G.\\
tunruh@frm2.tum.de
\end{center}

At the new neutron source FRM--II in Garching, Germany, the TACO
control system, originally developed at the ESRF in Grenoble, France is used for
instrument control purposes. TACO provides an object oriented,
distributed control system including a clearly defined API. In order to
equip TACO with a general user front end, a network based instrument
control system named NICOS has been developed at the FRM-II.
NICOS is divided into three parts: the NICOS Client, the NICOS Server and
the NICOSMethods. 

\vspace{3mm}
\underline{\bf 1. Introduction}

In this article a survey of the design of NICOS ({\bf n}etwork based {\bf i}nstrument
{\bf co}ntrol {\bf s}ystem) and a comprehensive
introduction on how to use NICOS is given.

In spite of the fact that the NICOS client/server, the NICOSMethods and TACO,
respectively, are separated program packages which do not depend on each other,
the whole software will be regarded in the following as one modular system for
control of complex instruments.

Besides the author of this article the following people at the FRM--II were
involved: J.~Beckmann, J.~Krüger, P.~Link, J.~Neuhaus and W.~Wein.

\vspace{3mm}
\underline{\bf 2. The NICOS client}

The NICOS client allows the user of an instrument to enter high level commands
or even command scripts to control complex operations, as e.g.\ 
a series of energy scans of a triple axis spectrometer. Python is used
as script language because it is easy to learn and provides powerful commands 
as well as a simple syntax which leads to short and well readable scripts.

The NICOS client itself is also written in Python (v.~2.1.1) using PyQt (v.~3.0.5)
as graphical library.

Almost every Python script can be executed via the NICOS client on the server. 
However at the moment there are some limitations:
\begin{itemize}
\item There is no interface for using the standard input stream. 
  Interactive communication between the Python script and the user is done via
  the special ``nicmd\_prg (prg)'' command which starts the program ``prg'',
  waits until it is finished and returns its stdout as Python string.
  The DISPLAY of the client which started the script is used as standard display
  for that program.
\item It is not safe to use special features as e.g.\ signal handlers, 
  multithreading and server functionalities. Multithreading is safe, however, when using the
  internal commands as e.g.\ the ``nicd\_startBgPrg'' function to start a process
  in a new thread.
\end{itemize}

The main window of the nicos client that appears after startup is displayed in
\begin{figure}[ht]
  \begin{center}
    \includegraphics[height=4cm]{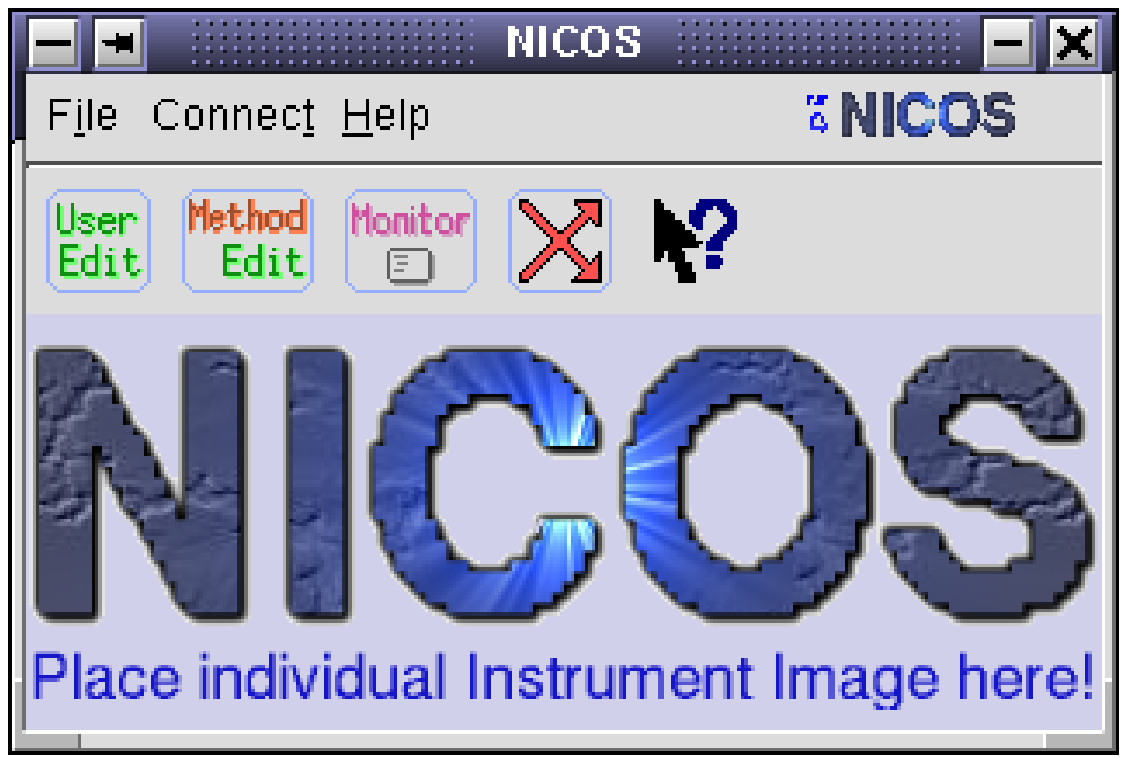}
    \CaptionFig{main}{}{The main window of the NICOS client}
  \end{center}
\end{figure}
Fig.~\ref{main}. After startup you have to enter the hostname and the port number 
(1199 by default) of the NICOS server you want to be connected to (connect menu).
Then you can login to that server with a user name and a password. After successful
login the arrows of the connect button in the main window switch from transparent
(grey) to red.

From the main window three major applications can be opened which will be described in
the following subsections: The user editor, the program control monitor and the 
configuration editor.

\vspace{3mm}
{\bf 2.1 The user editor}

The user editor window includes a simple text editor with common editing capabilities
\begin{figure}[ht]
  \begin{center}
    \includegraphics[height=12cm]{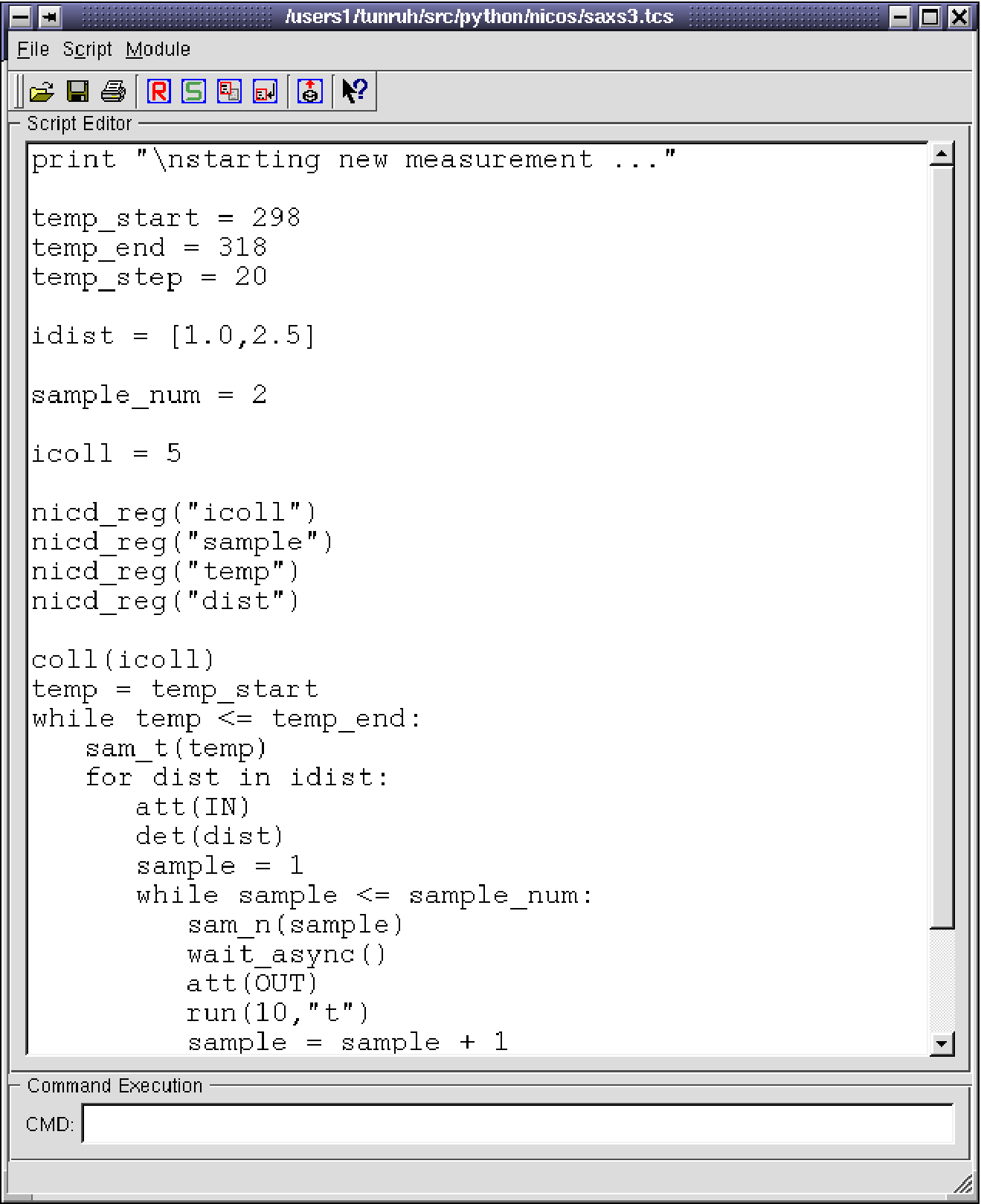}
    \CaptionFig{usereditor}{}{The user editor window of the NICOS client}
  \end{center}
\end{figure}
for the user to write experiment control scripts. Single line commands can be entered
using the botton command line with history functionality (cf. Fig.~\ref{usereditor}).
Via the menu or the buttons of the toolbar it is possible to execute a script on the
server (run) or to download the currently loaded script from the server to the editor.

It is possible to change the source code of a running script in the editor
at lines which have not yet been executed and which are not inside the logical
block of the current line (cf.\ subsection 2.2) and update the script with the new
code during runtime. 

Clicking on the reload tool button will recursively reload all
modules loaded by the server by default. These modules can be specified in the 
configuration file of the server. Thus it is possible to activate any change of
instrument configuration or of instrument control code by one mouse click.

The ``simulation'' button is not yet implemented, but will follow soon. Clicking
on this button will then also start the script on the server but in a simulation
mode. In this mode no hardware will be activated, but as many functionality checks
of the script as possible are performed to avoid e.g.\ syntax errors or out of 
limit errors occur when running the script in real mode.

\vspace{3mm}
{\bf 2.2 The program control monitor}

The window of the program control monitor is displayed in Fig.~\ref{monitor}.
\begin{figure}[ht]
  \begin{center}
    \includegraphics[height=14cm]{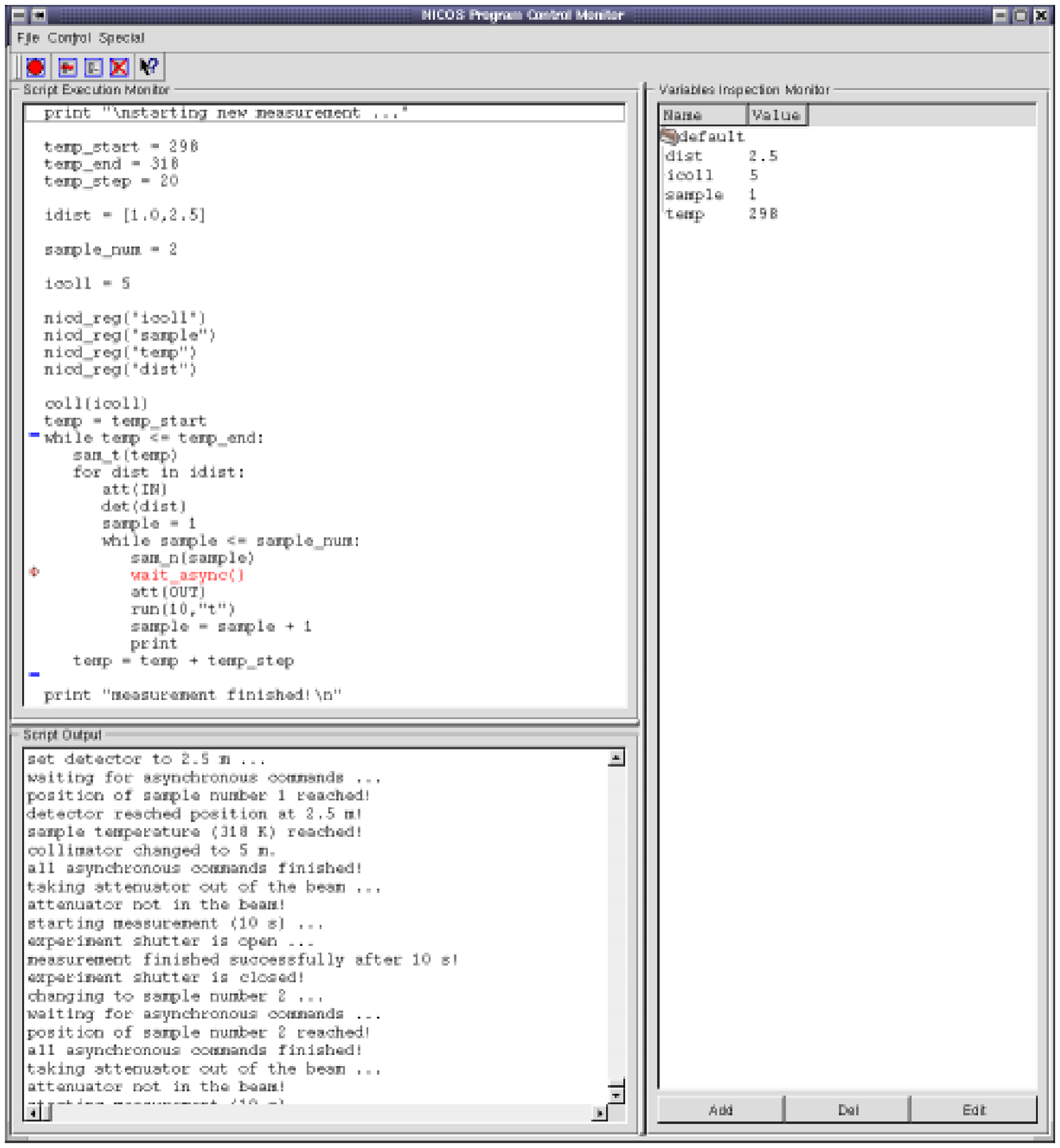}
    \CaptionFig{monitor}{}{The program control monitor window of the NICOS client}
  \end{center}
\end{figure}
Several informations about the status of the NICOS server and the
program script that is currently executed by the server are displayed. The window
is divided in three subwindows: the script execution monitor, the script output and
the variables inspection monitor.

A toolbar with buttons is positioned above the subwindows. The script execution 
status is indicated by the color of the leftest button: A red color means that
a script is currently executed, green means no script is running; yellow means 
that script execution is interrupted and transparent (grey) means that the client
is not connected to a server.

By means of the other three tool buttons a script can be interrupted, continued
or stopped by a simple mouse click.

In the {\bf script execution monitor} (upper left subwindow) the source code of the 
loaded script is displayed. The currently executed line is highlighted and marked
with an arrow on the left side. The logical block in which this line is located
is marked by blue lines. The update rate of this window can be adjusted.

The output (stdout and stderr) of the script is displayed in the {\bf script 
output window} (lower left subwindow). Python error messages are displayed here 
according to the standard Python shell.

In the {\bf variables inspection monitor} (right subwindow) variables and their
values can be displayed. It is possible to add and delete values from the script
using the ``nicd\_reg'' and ``nicd\_unreg'' commands. Addition, deletion and change
of entries are also possible during runtime pressing the ``Add'', ``Del'' and ``Edit''
buttons. Instead of variables it is also allowed to enter any expression. Double
clicking on an entry in the variables list causes the pop up of a window in which
the value of the selected variable can be changed.

Another possibility to influence the script execution during runtime is to use
the command line input window available from the special menu. Here any command
can be entered and executed in the context of the current script. 

The manipulation of instrument control scripts during runtime is intended to be
used e.g.\ to extend the range of an energy scan at a triple axis spectrometer
or to increase the measuring time at a TOF instrument, when the data collected
up to this moment indicate that these changes are useful. In this way beam time will be
saved because measurements or parts of measurements do not need to be performed
twice.

\vspace{3mm}
{\bf 2.3 The configuration editor}

The configuration editor has been developed at the ESRF in close contact with the 
\begin{figure}[ht]
  \begin{center}
    \includegraphics[height=10cm]{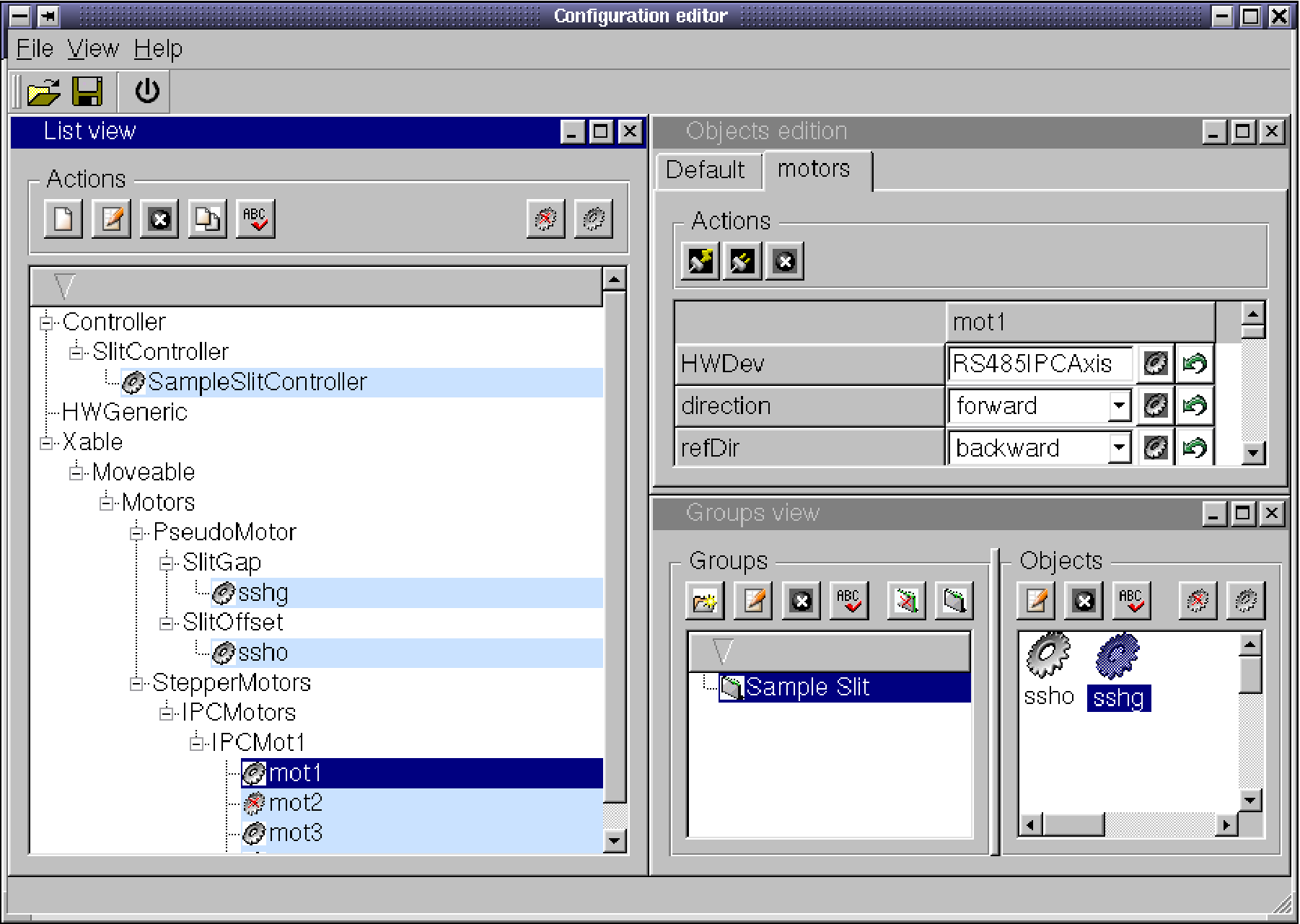}
    \CaptionFig{configeditor}{}{The configuration editor of the NICOS client}
  \end{center}
\end{figure}
FRM--II. The editor allows to create configuration files for python classes, which 
define e.g.\ components of an instrument. This tool has been developed to realize
the aim  {\bf ``configuring instead of programming''} claimed by the NICOSMethods.

The format of the configuration files will be converted automatically by the nicos
client to the standard format of the NICOSMethods which are described later in this
article. Thus all components of an instrument defined in the NICOS environment can
be configured by this editor. It is also possible to configure TACO servers if the
servers provide runtime configuration. 

\vspace{3mm}
\underline{\bf 3. The NICOS server}

The NICOS server is a multithreaded tcp stream server written in Python (v.\ 2.1.1).
\begin{table}[ht]
  \begin{center}
    \CaptionTab{servercmds}{}{List of all available commands of the NICOS server}
    \begin{tabular}{|l|l|}
      \hline
      {\bf command} & {\bf description}\\\hline
      get\_info & \parbox{12cm}{returns repr(nicd\_prg\_info). nicd\_prg\_info is a dictionary 
        which contains among other entries the loaded script text (prg\_text).}\\\hline
      get\_status & \parbox{12cm}{returns repr(nicd\_prg\_status). nicd\_prg\_status is a dictionary 
        which contains status information of the server e.g.:\ modules available at 
        startup (modules)}\\\hline
      get\_control\_info & \parbox{12cm}{returns repr(nicd\_prg\_control). nicd\_prg\_control is a dictionary 
        which contains some program control information e.g.: priority of owner (from passwd file)
        (priority), user\_name (login) of the owner (user\_name)}\\\hline
      get\_values & \parbox{12cm}{returns a dictionary (as repr(nicd\_x)) which contains names and
        values of all registered items (cf.\ section 2)}\\\hline
      get\_value (name) & \parbox{12cm}{returns the value of the variable ``name'' in the 
        current namespace of a running prg}\\\hline
      set\_values (idict) & \parbox{12cm}{sets items found in idict.keys() to the 
        specified expressions} \\\hline
      add\_values (ilist) & \parbox{12cm}{adds items in ``ilist'' to nicd\_x (registers items, 
        cf.\ section 2)} \\\hline
      del\_values (ilist) & \parbox{12cm}{deletes items in ``ilist'' from nicd\_x (unregisters items, 
        cf.\ section 2)} \\\hline
      reload\_modules & \parbox{12cm}{performs recursive reload of modules loaded by default} \\\hline
      exec\_cmd (prg) & \parbox{12cm}{executes a python command in the current namespace of a running prg} \\\hline
      config  & \parbox{12cm}{starts the configuration editor} \\\hline
      update\_prg (prg) & \parbox{12cm}{updates the python script with prg code found in the 
        ``prg'' string} \\\hline
      break\_prg & \parbox{12cm}{interrupts prg execution} \\\hline
      cont\_prg & \parbox{12cm}{continues an interrupted prg} \\\hline
      stop\_prg & \parbox{12cm}{stops an interrupted prg} \\\hline
      release\_control & \parbox{12cm}{releases the control of prg, this means that a program started by 
        a user with a certain priority can be influenced by another user which has a lower priority} \\\hline
      get\_prg & \parbox{12cm}{returns source code of current prg} \\\hline
      get\_output & \parbox{12cm}{returns output of prg since last call of this function} \\\hline
      \_get\_output & \parbox{12cm}{returns output of prg since prg startup} \\\hline
      exit & \parbox{12cm}{closes connection to server} \\\hline
      \_exit & \parbox{12cm}{closes connection to server without release of prg control (cf.\ release command)} \\\hline
    \end{tabular}
  \end{center}
\end{table}
It provides multiple user login and a secure password authentification procedure.
Every login is associated with a security level number, which can be determined by
a script that runs on the server. Thus it is possible to implement execution 
permissions for users within the NICOSMethods (cf.\ below).

The server can be configured by a simple configuration file. In this file the
Python modules which implement the whole instrument can be specified. Executing
the first script will initiate the import of these modules.
To reload the modules recursively a deep\_reload function of the server is available
and can easily be activated using the reload button of the NICOS client.
User names can be added in the configuration file and passwords can be set and changed
using the ``nicd\_passwd.py'' script.

Python scripts started on the server will be divided into blocks, which are executed 
sequentially. Before execution a syntax check of the whole script is performed.

By using Python special client classes or a generic Python TACO client class (as
provided by the ESRF), it is possible to communicate with TACO device servers directly.

The capabilities of the NICOS server are described in section~2 of this 
article. Comprehensive information for all server commands available so far are listed in 
Tab.~\ref{servercmds}.

\vspace{3mm}
\underline{\bf 4. The NICOSMethods}

NICOSMethods are Python classes which represent standard devices of instruments 
such as motors, digital i/o cards and counters. There are three important advantages
to introduce such standard devices: 

\begin{enumerate}
\item Some general functionalities as e.g.\ 
  grouping (cf.\ below) can be implemented in the basic classes and therefore it needs
  not to be written for each special device. 
\item It is possible to define a clear interface for important classes of devices. Such
  an interface allows to write powerful global commands (e.g.\ a scan command) which 
  can be used for all the special devices in this class. 
\item The interfaces include standard functions for the configuration of the devices.
  This is needed to set up a general configuration tool as the configuration editor
  of the NICOS client, which can also be used for any device implemented in the future.
\end{enumerate}

The NICOSMethods framework is written in a way that the code for a new device
\begin{table}[ht]
  \begin{center}
    \CaptionTab{xable}{}{List of the commands of the standard interface of an Xable object}
    \begin{tabular}{|l|l|}
      \hline
      {\bf command} & {\bf description}\\\hline
       start (value) & \parbox{12cm}{starts object's main operation (OMO)}\\\hline
       read () & \parbox{12cm}{returns current value of the MPSD of the device}\\\hline
       status() & \parbox{12cm}{returns current status of the device (log int, bit array)}\\\hline
       setPar (key,value) & \parbox{12cm}{sets configuration parameter ``key'' to ``value''}\\\hline
       getPar (key) & \parbox{12cm}{returns current value of the configuration parameter ``key''}\\\hline
       init () & \parbox{12cm}{initializes the device}\\\hline
       wait (timeout) & \parbox{12cm}{waits until OMO is finished; if timeout is specified 
         this parameter overwrites a default timeout}\\\hline
       stop () & \parbox{12cm}{stops OMO}\\\hline
       abort () & \parbox{12cm}{aborts OMO (emergency halt)}\\\hline
       reset () & \parbox{12cm}{resets the device to a clearly defined state}\\\hline
    \end{tabular}
  \end{center}
\end{table}
is easy to write, short and clear.

The root class for all devices is a class called Xable. An Xable object represents 
a physical or a virtual device, which can be abstracted to control or measure a
single physical state of an object (main physical state, MPS), which is represented 
by the main physical state descriptor (MPSD) such as a position (length, angle), an 
energy, a temperature, the position of a switch, a count rate or a 2-dimensional 
intensity distribution. The MPSD can be read and/or set by the main operation of
an object (OMO).

In this class the standard interface of devices is defined. The interface definition 
is described in Tab.~\ref{xable}. It is intended to add a pause~() and a getAllPars~()
command.

There are several parameters defined for an Xable object. The ``name'' parameter can be used
to specify an alias of the instance name of an Xable object. The ``adev'' dictionary
specifies attached devices which are selected by configuration. These devices are
linked to hard coded internal names of the Xable object.

With the standard ``controller'' parameter a device can specify the instance name of 
its controller object. Such objects are called ``exported objects'' of the controller.
Several different devices can have the same controller. If a 
controller is specified, the functions of the standard interface are expected to be 
defined in the controller object. From this point of view a controller offers the
possibility to have the same code for different devices. The pointer to the Xable
object that calls a method of the controller class will be added to the argument list.
Thus, the controller can determine from which device a function call originates.

The controller together with its exported devices is called ``component''.

The controller concept reflects the possibility of TACO servers to export several
devices. Attached devices correspond to the possibility of TACO servers to communicate
with other TACO servers. So it is evident that the only general difference between 
TACO servers and NICOS Xable objects is that the Xable objects have no server
functionality. Recently a project at FRM--II was initiated with the aim to write
a generic TACO server which can add the server capabilities to any Xable object.
Then the NICOSMethods provide another way to develop TACO servers: 

\centerline{\bf Build a server by writing a client!}

This concept is of special interest since it is very easy to implement
a device funtionality with the script language Python and the NICOSMethods. The only
thing to do is to write a class which inherits the Xable class. In this class for each
standard interface function a method named ``do'' + $<$name of the interface 
function$>$\ must be written. The following code example implements a class MotorWithSwitch which
has two attached devices with the internal names ``mot1'' and ``switch1''. mot1
is a moveable object. A moveable object is an Xable object that defines the alias 
``move'' of the ``start'' function of the standard interface. switch1 is a 
switchable object. A switchable object is an Xable object that defines the alias 
``switchTo'' of the ``start'' function of the standard interface. The MotorWithSwitch
class inherits from the class moveable. It represents the functionality of
mot1 with the exception that it only starts the mot1 if switch1 is set
to ``ON''. 

\begin{verbatim}
from nicm_def import *
class MotorWithSwitch (Moveable):
    typelist = {
        "mot1":   Moveable,
        "switch1":  Switchable
    }
    def doStatus (self): 
        return self.mot1.read ()
    def doStart (self, position):
        if self.switch1.read() == "ON":
            self.mot1.start()
        else:
            NicmError ("switch1 not on")
    def doRead (self): 
        return self.mot1.read ()
    def doWait (self): 
        self.mot1.wait ()
\end{verbatim}

Another feature of the NICOSMethods framework is the grouping of Xable objects.
To create a group ``g'' of the Xable objects ``a'', ``b'' and ``c'', \verb/g = XOGroup(a,b,c)/
has to be written. The command \verb/g.read ()/ returns a tuple of the 
return values of the read commands of the objects ``a'', ``b'' and ``c''. This
works for all common functions.

An important feature of the grouping concept is that a particular controller receives
only a single function call even if more than one of the grouped objects are
exported devices of the controller. Accordingly the standard interface functions
of a controller expect as arguments a list of all devices and a corresponding list of 
function arguments specified in the group call. The controller methods have to return 
a tuple with a corresponding number of return values.

Therefore it is possible to implement devices with dependent functionalities which 
can be handled simultaneously in the controller and initiated with a single function call.

Because of the limited space it was only possible to describe the most important features
of the NICOSMethods. 

The NICOSMethods are under development since december 2001. Nevertheless up to now
the control software of a three axis spectrometer at the FRM--II using TACO and the 
NICOSMethods is nearly completed. Another instrument started software development 
with the NICOSMethods and further instruments intend to use this software package.
The response of the NICOS users is very positive and as was demanded the development
will be continued.
\end{document}